\documentclass{ws-ijmpd}

\usepackage{graphicx}
\usepackage{multirow}

\renewcommand{\vec}[1]{{\bf{#1}}}

\begin{document}

\markboth{}{Independent analysis of the orbits of Pioneer 10 and 11}

\title{\uppercase{Independent analysis of the orbits of Pioneer 10 and 11}}

\author{\uppercase{Viktor T. Toth}\footnote{\texttt{http://www.vttoth.com/}}}

\maketitle

\begin{abstract}
Independently developed orbit determination software is used to analyze the orbits of Pioneer 10 and 11 using Doppler data. The analysis takes into account the gravitational fields of the Sun and planets using the latest JPL ephemerides, accurate station locations, signal propagation delays (e.g., the Shapiro delay, atmospheric effects), the spacecrafts' spin, and maneuvers. New to this analysis is the ability to utilize telemetry data for spin, maneuvers, and other on-board systematic effects. Using data that was analyzed in prior JPL studies\cite{JPL1998,JPL2002}, the anomalous acceleration of the two spacecraft is confirmed. We are also able to put limits on any secondary acceleration (i.e., jerk) terms. The tools that were developed will be used in the upcoming analysis of recently recovered Pioneer 10 and 11 Doppler data files.
\end{abstract}


\section{Introduction}

As first reported in 1998\cite{JPL1998}, and later confirmed in 2002\cite{JPL2002}, the Pioneer 10 and 11 spacecraft have experienced a small, approximately constant, sunward-pointing acceleration during the later interplanetary cruise phase of their missions. This acceleration is approximately $8.74\times 10^{-10}$~m/s$^2$ in magnitude; within the achievable margin of error, its value remains constant during the time period spanned by the data sets (11.5 years for Pioneer 10, 3.75 years for Pioneer 11) that were investigated.

Although a satisfactory explanation is yet to be found, the authors of the two studies\cite{JPL1998,JPL2002} stressed that the most likely cause of the anomalous acceleration is an unaccounted for on-board systematic force, such as a fuel leak, outgassing, or anisotropic thermal radiation. Although an attempt was made to account for these forces, due to the lack of available information no definitive conclusions were reached.

Recently\cite{MDR2005} we successfully recovered the entire mission record of Pioneers 10 and 11 in the form of telemetry data files containing Master Data Records (MDRs.) The ultimate goal of the effort described in this paper is to develop orbital analysis tools that can estimate the thermal emissions from the spacecraft and the associated recoil force using telemetry. Before that effort can proceed, however, it is necessary to validate our tools by re-analyzing previously studied data sets.

We begin this paper by briefly describing the Pioneer spacecraft and missions in Section~\ref{sec:pioneer}. In Section~\ref{sec:orbit}, we describe an independently developed approach for orbit determination. Section~\ref{sec:results} presents results: the anomalous acceleration is confirmed, validating the tools presented here and demonstrating that they can be used to achieve sufficient precision to study the Pioneer spacecraft. We also characterize the extent to which a jerk term may be present in the Doppler data. Lastly, Section~\ref{sec:plans} describes our plans to use these tools and related expertise to explore recently recovered Doppler and telemetry data, the goal being a satisfactory and complete description of the twin Pioneer missions in their entirety, from launch (or earliest Doppler data point available) until the end of mission.

\section{The Pioneer 10 and 11 spacecraft}
\label{sec:pioneer}

\begin{figure*}
\centering
\includegraphics[width=\linewidth]{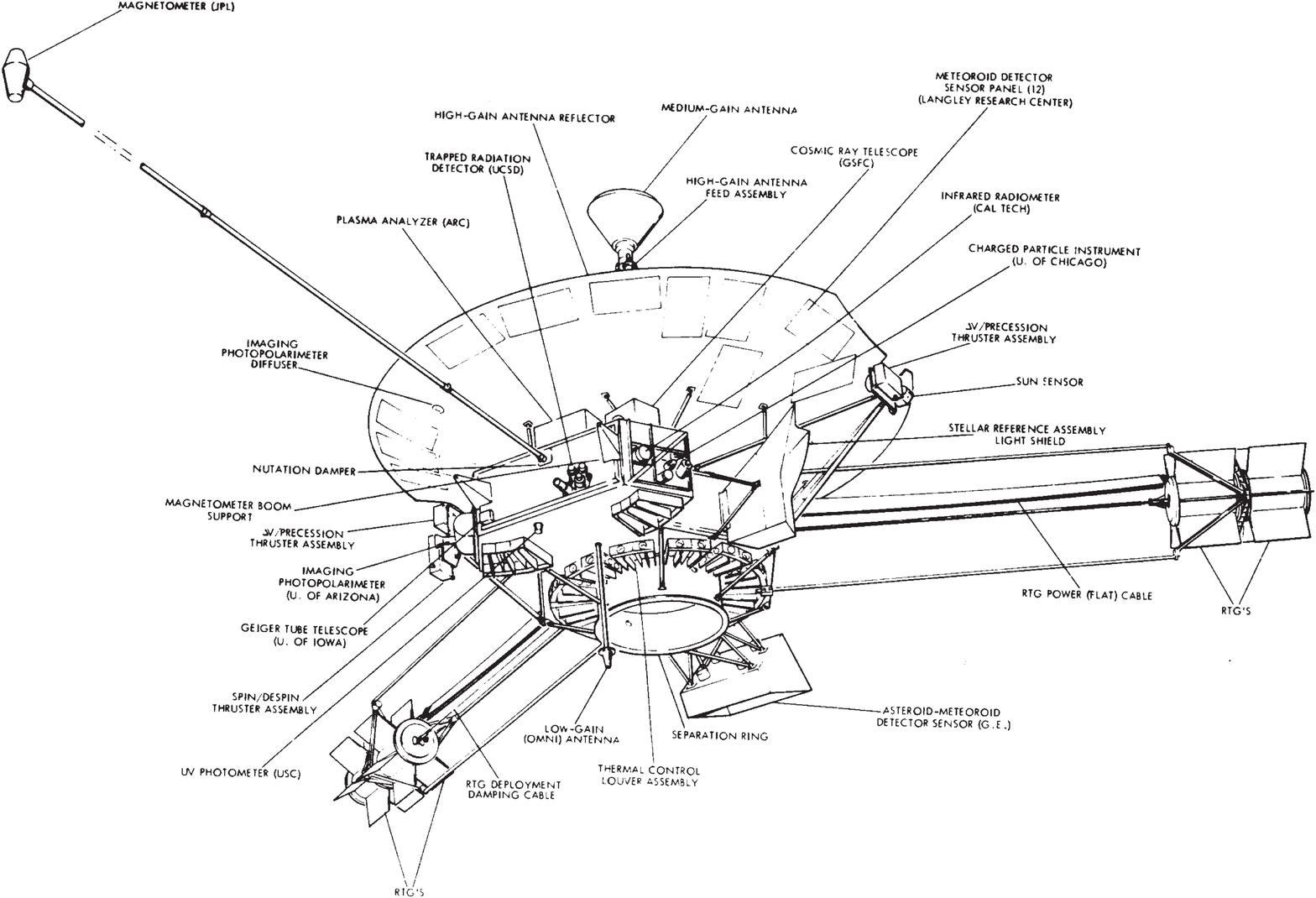}
\caption{The Pioneer spacecraft.}
\label{fig:pioneer}
\end{figure*}

\subsection{The Pioneer spacecraft}

Launched in 1972 (Pioneer 10) and 1973 (Pioneer 11), these two Pioneer spacecraft were the first to traverse the asteroid belt, fly by Jupiter, Saturn (Pioneer 11 only), and explore the outer solar system. Originally designed to operate for at least 18 months, the spacecraft far exceeded expectations: Pioneer 11 remained functional until 1995, whereas Pioneer 10 sent its last signal more than 30 years after its launch, in 2003.

These two Pioneer spacecraft remain the most precisely navigated deep space vehicles to date. This is due to several factors, most notable among them is the fact that the spacecraft are spin stabilized: therefore, they require relatively few attitude correction maneuvers, which means that most of the time, they fly undisturbed, their trajectories governed solely by the laws of celestial mechanics.

In appearance (Figure~\ref{fig:pioneer}), the Pioneer spacecraft are dominated by a 2.74 m high gain antenna (HGA). The spacecraft body is situated behind the HGA, and consists of a main compartment housing most spacecraft systems, and a science compartment that contains science instruments. Two approximately 3~m long booms each hold a pair of radioisotope thermoelectric generators (RTGs) (4 in total). A third boom, approximately 6 meters in length, holds a magnetometer. The total (wet) mass of the Pioneer spacecraft is $\sim$260 kg, of which $\sim$30 kg is propellant and pressurant.

The main body of the spacecraft is well insulated, designed to maintain acceptable operating temperatures for all on-board systems both in the near-Earth environment, exposed to the Sun, and also far from the Sun in deep space. A passive louver system with springs and bimetallic actuators, radiatively coupled to the spacecrafts' electronics platforms, ensured that excess heat was radiated away in space when the spacecraft was warm, but also that heat was retained when the spacecraft was far from the Sun.

The Pioneer spacecraft communicated with the Deep Space Network using S-band microwave transmissions. The nominal uplink frequency was approximately 2.1 GHz, the nominal downlink frequency 2.29 GHz. The spacecraft provided no range observable. The spacecraft also lacked a precision oscillator on board; however, their transmitter could operate in coherent mode. In this mode, the transmitter locked on to the frequency of the received signal, and the transmitted signal had a frequency that was an exact rational multiple (240/221) of the received signal's frequency. In this mode, the accuracy of the received signal was limited only by the stability of the ground-based transmitter, which allowed precision Doppler observations to be made.

\subsection{The Pioneer anomaly}

The Pioneer anomaly is best described as our apparent inability to fit a model based on the known laws of celestial mechanics to the Pioneer Doppler observations. The procedure can be briefly described as follows: a set of equations is used to describe the motion of the Pioneer spacecraft in the solar system, relative to the participating stations of the Deep Space Network. The motion of the spacecraft is determined by the initial state vector (three position and three velocity components) and perhaps additional parameters, such as values that characterize the velocity change as a result of maneuvers. Using an optimization technique such as the nonlinear least squares method, best fit values for the initial state vector and additional parameters are obtained, minimizing the difference between calculated and observed Doppler values.

When the model is modified to include an additional fictitious force that results in a constant sun-pointing acceleration of the spacecraft, the statistical fit between the model and observations is vastly improved\cite{JPL2002,CBM2005}. For this reason, it has been assumed that a force of unknown origin, resulting in an approximately constant, approximately sunward acceleration of $\sim 8.74\times 10^{-10}$~m/s$^2$, acts on the spacecraft.

Since the initial discovery of this discrepancy between computed and observed Doppler readings, many attempts have been made to explain the anomaly using both conventional and new physics. To date, none of these attempts produced a satisfactory result. In particular, the possibility that the anomalous result is due partially or entirely to on-board forces could not be excluded; however, insufficient data was available to properly characterize on-board forces, resulting in wildly differing estimates, including assertions that on-board forces are far from sufficient to explain the anomalous acceleration\cite{JPL1998,JPL2002} to suggestions that the anomalous acceleration is due entirely to on-board forces\cite{KATZ1998,MURPHY1998,LKS2003}.

\section{Orbit determination}
\label{sec:orbit}

The data that were used to develop the most precise estimate to date of the anomalous acceleration of Pioneer 10/11\cite{JPL2002} have been made available to interested researchers. Since the anomalous acceleration has been confirmed by several independent teams, it is reasonable to expect that any new attempt at orbit determination should be able to reproduce these well established results. The anomalous acceleration, therefore, serves as a useful test case to validate orbit determination software.

Our orbit determination is based on well established methods and techniques\cite{MOYER2000,MG2005}. It can be briefly described as follows: we integrate with high precision the (optionally relativistic) equations of motion for the Pioneer spacecraft in the gravitational field defined by major solar system bodies. The position of solar system bodies, along with the positions of ground-based Deep Space Network stations, are obtained from a software library provided by JPL. We compute the Doppler shift of the radio signal sent to, and received from, the spacecraft, applying all necessary corrections.

\subsection{Timekeeping}

Equations of motion are integrated using Ephemeris Time (ET). Converting from ET to atomic time (TAI) can be accomplished using the following approximate formula (see Eqs. 2-26 through 2-28 of Ref.~\refcite{MOYER2000}):
\begin{equation}
\mathrm{ET}-\mathrm{TAI}=32.184+1.657\times 10^{-3}\sin E,
\end{equation}
where
\begin{eqnarray}
E&=&M+0.01671\sin M,\\
M&=&6.239996+1.99096871\times 10^{-7}t,
\end{eqnarray}
and $t$ is ET in seconds past J2000.

Data files from tracking stations are using UTC timestamps. The difference between UTC and TAI is an integer number of seconds ranging from 10 to 32 as leap seconds were added (UTC is a discontinuous time scale.)

\subsection{Equations of motion}

To calculate the orbit of the Pioneer probes, we used the Newtonian equations of motion without relativistic corrections. In the outer solar system, this provides a solution with adequate accuracy. For a spacecraft moving in the solar system that consists of $n$ bodies, the spacecraft's motion is governed by the equation
\begin{equation}
\ddot{\vec{r}}=\sum_{i=1}^n\mu_i\frac{\vec{r}_i-\vec{r}}{|\vec{r}_i-\vec{r}|^3},
\label{eq:motion}
\end{equation}
where $r$ is the spacecraft's position vector, $\mu_i$ is the mass of the $i$-th solar system body, multiplied by the gravitational constant, and $\vec{r}_i$ is the position vector of the $i$-th solar system body, while the dot denotes differentiation with respect to time.

We carry out precision integration of ($\ref{eq:motion}$) using Stoermer's rule and Richardson's extrapolation technique. The integrator was validated by computing the orbits of various solar system bodies and comparing the results against data from the JPL HORIZONS database\footnote{See {\tt http://ssd.jpl.nasa.gov/?horizons}.}.

\subsection{Relativistic equations of motion}

In the presence of a strong gravitational field or when velocities are high, the equations of motion (\ref{eq:motion}) must be modified, taking into account effects due to general relativity. Equation 4-26 in Ref.~\refcite{MOYER2000} takes into account all effects up to $(v/c)^2$ using the parametric post-Newtonian (PPN) formulation of general relativistic theories. Moyer's formulation expresses relativistic corrections in terms of the positions, velocities, and accelerations of interacting bodies. For Pioneer, we can safely omit terms that contain accelerations. We can also drop terms containing factors of $1/|\vec{r}_j-\vec{r}_k|$ where $j\ne k$ are indices representing solar system bodies; these terms represent nonlinear contributions of pairs of distant bodies, and do not significantly affect the Pioneer orbits. After some trivial algebra, we can express the resulting relativistic equations of motion in the form
\begin{equation}
\ddot{\vec{r}}=\sum_{i=1}^n\frac{\mu_i[(\vec{r}_i-\vec{r})A_i+\vec{B}_i]}{|\vec{r}_i-\vec{r}|^3} ,
\end{equation}
where $A_i$ and $\vec{B}_i$, which are themselves functions of the positions and velocities of the gravitating bodies involved and the test particle, can be written as
\begin{equation}
A_i=1-\frac{1}{c^2}\Bigg\{\frac{2(\beta+\gamma)}{|\vec{r}_j-\vec{r}|}\sum\limits_{j=1}^n\mu_j+\gamma \dot{\vec{r}}^2+(1+\gamma)\dot{\vec{r}}_i^2
-2(1+\gamma)\dot{\vec{r}}\cdot\dot{\vec{r}}_i-\frac{3}{2}\left[\frac{(\vec{r}-\vec{r}_i)\cdot\dot{\vec{r}}_i}{|\vec{r}_i-\vec{r}|}\right]^2\Bigg\},~~~
\end{equation}
and
\begin{equation}
\vec{B}_i=\frac{1}{c^2}\left\{(\vec{r}-\vec{r}_i)\cdot[(2+2\gamma)\dot{\vec{r}}-(1+2\gamma)\dot{\vec{r}}_i]\right\}(\dot{\vec{r}}-\dot{\vec{r}}_i).
\end{equation}

In the non-relativistic limit ($c\rightarrow\infty$), $A_i=1$ and $\vec{B}_i=0$, and we get back the Newtonian equations of motion.

The parameters $\beta$ and $\gamma$ are parameters of the post-Newtonian formulation; for Einsteinian relativity, they both have the value of 1.

\subsection{Oblateness}

Oblateness of a planetary body modifies its gravitational field. The difference between the actual potential field of an oblate body and a spherically symmetric field can be described using spherical harmonics.

The Pioneer spacecraft had close encounters with Jupiter and (for Pioneer 11 only) Saturn. Both of these planetary bodies are rotationally symmetrical. Their gravitational potential field due to body $i$ is described by the following equation:
\begin{equation}
U_i=\frac{\mu_i}{|\vec{r}_i-\vec{r}|}\left[1-\sum_{k=1}^\infty\frac{J_k^{(i)}a_i^kP_k(\sin\theta)}{|\vec{r}_i-\vec{r}|^k}\right],
\label{eq:Ui}
\end{equation}
where $P_k(x)$ is the $k$-th Legendre polynomial in $x$, $a_i$ is the equatorial radius of planet $i$, $\theta$ is the latitude of the spacecraft relative to the planet's equator, and $J_k^{(i)}$ is the $k$-th spherical harmonic coefficient of planet $i$.

In order to put (\ref{eq:Ui}) to use, first it must be translated into an expression for force by calculating its gradient. Second, it is also necessary to express the position of the spacecraft in a coordinate system that is fixed to the planet's center and equator.

In a Cartesian coordinate system with the $x_1x_2$ plane fixed to the planet's equator, we have
\begin{eqnarray}
r=|\vec{r}-\vec{r}_i|&=&\sqrt{x_1^2+x_2^2+x_3^2},\\
\theta&=&\arctan\frac{x_3}{\sqrt{x_1^2+x_2^2}}.
\end{eqnarray}

The gradient of the potential field can be expressed as
\begin{equation}
\frac{\partial U_i}{\partial x_j}=\frac{\partial U_i}{\partial r}\frac{\partial r}{\partial x_j}+\frac{\partial U_i}{\partial\theta}\frac{\partial\theta}{\partial x_j}.
\end{equation}

Individual partial derivatives can be calculated as
\begin{eqnarray}
\frac{\partial r}{\partial x_j}&=&\frac{x_j}{r},\\
\frac{\partial\theta}{\partial x_j}&=&-\frac{x_jx_3}{r^2\sqrt{(x_1^2+x_2^2)}},\\
\frac{\partial\theta}{\partial x_3}&=&\frac{\sqrt{x_1^2+x_2^2}}{r^2},\\
\frac{\partial U_i}{\partial r}&=&-\frac{\mu_i}{r^2}+\mu_i\sum_{k=2}^\infty\frac{(k+1)J_k^{(i)}a_i^kP_k(\sin\theta)}{r^{k+2}},\\
\frac{\partial U_i}{\partial\theta}&=&-\mu_i\sum_{k=2}^\infty\frac{J_k^{(i)}a_i^kP'_k(\sin\theta)\cos\theta}{r^{k+1}},
\end{eqnarray}
where $P'_k(x)=dP_k(x)/dx$.

To use these formulae, the position of the spacecraft must be expressed in a coordinate system that is rotated to coincide with the planet's equatorial plane. The direction of the planet's North pole (which is normal to the equatorial plane) is known in the form of its right ascension ($\alpha$) and declination ($\delta$) angles.

To express the coordinates of a spacecraft using a frame of reference with its origin at the center of planet $i$, one only needs to calculate $\vec{r}-\vec{r}_i$. To express this in a frame of reference rotated by $\alpha$ in the $x_1x_2$ plane, we have
\begin{eqnarray}
x'_1&=&x_1\cos\alpha+x_2\sin\alpha,\\
x'_2&=&-x_1\sin\alpha+x_2\cos\alpha,\\
x'_3&=&x_3.
\end{eqnarray}

Finally, to express this in a frame of reference further rotated by $\delta$ in the $x'_1x'_3$ plane, we have
\begin{eqnarray}
x''_1&=&x'_1\sin\delta-x'_3\cos\delta,\\
x''_2&=&x'_2,\\
x''_3&=&x'_1\cos\delta+x'_3\sin\delta.
\end{eqnarray}

These two transformations can be combined into a single transformation matrix that can be used to translate coordinates to an equatorial coordinate system; the inverse of this matrix can be used to translate any resulting forces back into the original frame of reference.

\subsection{Time delay}

Our observable is radio-metric data, measured at specific moments in time at specific Earth stations. To accurately compute a matching orbit, we need to know not where the spacecraft was at the moment the signal was received on the Earth, but where the spacecraft was at the moment it transmitted that signal.

The time delay can be calculated using an iterative process that is known to converge rapidly. We assume that the signal propagates from the transmitter at $\vec{P}_1(t)$ to the receiver at $\vec{P}_2(t)$. If the time of reception $t_2$ is known, the location of the receiver can be accurately calculated as $\vec{P}_2(t_2)$. We use $t_2$ as a first crude estimate for the time of transmission:
\begin{equation}
t_{1(0)}=t_2.
\end{equation}
The corresponding location along the transmitter's path is $\vec{P}_1(t_{1(0)})$, which we use to estimate the propagation delay and arrive at an improved estimate of the time of transmission: $t_{1(1)}=t_2-|\vec{P}_1(t_{1(0)})-\vec{P}_2(t_2)|/c$. This process can be improved iteratively, using
\begin{equation}
t_{1(i)}=t_2-\frac{1}{c}|\vec{P}_1(t_{1(i-1)})-\vec{P}_2(t_2)|.
\end{equation}

By itself, this procedure alone does not provide a sufficiently accurate estimate for the time delay. The reason is that the radio signal of the Pioneer probes travels through the gravitational field of the Sun and the planets, where the effective speed of light is less than $c$. The consequent delay $\Delta t$, known as the Shapiro time delay, between two points $\vec{P}_1$ and $\vec{P}_2$ can be calculated as (cf. Ref.~\refcite{MOYER2000} Eq. 8-54):
\begin{equation}
\Delta t=\frac{l}{c}\ln\frac{r_1+r_2+r_{12}+l}{r_1+r_2-r_{12}+l},
\end{equation}
where $l=\mu(1+\gamma)/c^2$; $\mu$ is the mass of the body whose potential field is being considered, multiplied by the gravitational constant; $\gamma$ is a parameter of the PPN (Parameterized Post-Newtonian) formalism that is equal to 1 for general relativity; $r_1$ and $r_2$ are the distances to $\vec{P}_1$ and $\vec{P}_2$ from the gravitating body; and $r_{12}=|\vec{P}_1-\vec{P}_2|$ is the distance between $\vec{P}_1$ and $\vec{P}_2$.

For a spacecraft moving in deep space far from planetary bodies, only the Sun's gravitational potential needs to be considered for the Shapiro delay. Furthermore, the distance of the spacecraft from the solar system barycenter can be substituted in place of the distance from the Sun's center of gravity, resulting in a substantial simplification of the calculations at no cost in terms of accuracy.

\subsection{Calculating the Doppler observable}
\label{sec:Doppler}

Our observable is Doppler: that is to say, the goal of the orbit determination exercise is to find an initial state vector that minimizes the difference between the measured frequency of the spacecraft's radio transmissions vs. the calculated frequency. As the spacecraft lacked a precision oscillator on board, this always means two-way or three-way Doppler: a signal of known frequency is transmitted to the spacecraft, which in turn reradiates a coherent response signal to the Earth, that is then received by either the same antenna that was used to do the transmission, or a different antenna.

If we know the exact positions and velocities of the transmitting and receiving stations as well as the spacecraft itself, the frequency of the received signal can be calculated to high precision. In an inertial coordinate system, with the effects of special relativity taken into account, a signal of frequency $f_1$ transmitted from station $\vec{P}_1$ (moving with velocity $\vec{v}_1$) will be received at station $\vec{P}_2$ (moving with velocity $\vec{v}_2$) at the frequency
\begin{equation}
f_2=f_1\frac{(c-v'_2)\sqrt{c^2-v_1^2}}{(c-v'_1)\sqrt{c^2-v_2^2}},\label{eq:Doppler}
\end{equation}
where $v'_1$ and $v'_2$ are the velocity components along the line $\vec{P}_1\vec{P}_2$, i.e.,
\begin{equation}
v'_i=\vec{v}_i\cdot\frac{\vec{P}_2-\vec{P}_1}{|\vec{P}_2-\vec{P}_1|},~~~~~(i=1,2).
\end{equation}

Doppler is not an instantaneous measurement: the receiving station counts the difference in cycles between the received frequency and the reference frequency for a specific interval. When the spacecraft is in deep space and the forces acting on it are very small, we can use the spacecraft's average velocity during the Doppler interval in (\ref{eq:Doppler}). However, this can lead to significant errors when the spacecraft is subjected to significant acceleration, e.g., when it is near a major solar system body.

Note that for the Earth stations, their instantaneous velocity must be accounted for properly, as the Earth stations are always accelerating as a result of the Earth's rotation and orbital motion.

\subsection{Nongravitational forces and effects}

In addition to gravity, several small forces act on the spacecraft and/or the transmitted radio signal that must be taken into account.

\subsubsection{Solar pressure}

Light from the Sun produces a noticeable amount of force on the spacecraft. Near the Earth, every square meter of surface receives 1366~W of solar power (``solar constant''); the total area of the Pioneer HGA is approximately 5.91~m$^2$, translating into more than 8~kW! (In contrast, the anomalous acceleration of the Pioneers is equivalent to $\sim$65~W of power emitted in a collimated beam in the appropriate direction.) Far from the Sun, solar pressure is significantly lower, but even at 20~AU, the HGA receives $\sim$20~W of sunlight, and therefore this effect cannot be ignored.

Since most of the time, the HGA is pointed towards the Earth and, consequently, in the approximate direction of the Sun, we need not model the spacecraft body behind the HGA; instead, we can use a ``flat disk'' model of the HGA. The HGA surface area is known. Sunlight that reaches the surface will transfer its momentum to the spacecraft; in addition, sunlight that is reflected off the surface, or reemitted by the surface, transfers additional momentum.

It is obvious that three principal directions need to be considered. Intercepted light will transfer momentum that is in the direction of the Sun-spacecraft line. Diffusely reflected, as well as re-emitted solar energy will produce momentum along the spin axis. Lastly, specularly reflected sunlight will produce momentum along a line that is the reflection of the Sun-spacecraft line along the spin axis, in the Sun-spacecraft-spin axis plane.

Quantitatively, force acting on the spacecraft due to intercepted sunlight can be calculated as
\begin{equation}
\vec{F}_\mathrm{intcpt}=\frac{sA(\vec{r}\cdot\vec{n})}{cr^4}\vec{r},
\end{equation}
where $s$ is the normalized solar constant ($s=1366$~W/m$^2\times$(1~AU)$^2$), $A=5.91$~m$^2$ is the surface area of the HGA, $\vec{n}$ is a normal vector in the direction of the spin axis pointing towards the rear of the spacecraft, and $\vec{r}$ is the Sun-spacecraft vector. As the value of $A$ is somewhat uncertain (the antenna is not exactly round, it has cutouts, and parts of it are shadowed by structural elements and the medium gain antenna) we may multiply this expression with a proportionality factor $\alpha\simeq 1$:
\begin{equation}
\vec{F}_\mathrm{intcpt}=\alpha\frac{sA(\vec{r}\cdot\vec{n})}{cr^4}\vec{r},
\end{equation}

The amount of specularly reflected sunlight will be proportional to the amount of intercepted sunlight, with a proportionality factor $\sigma$. Its direction can be computed using elementary vector geometry:
\begin{equation}
\vec{F}_\mathrm{spec}=-\sigma\left[\vec{F}_\mathrm{intcpt}-2(\vec{F}_\mathrm{intcpt}\cdot\vec{n})\vec{n}\right].
\end{equation}

Whatever solar radiation was not specularly reflected is either reflected diffusely, or it is absorbed by the HGA and reradiated. Either way, the resulting force will be in the direction of the spin axis. The force will be proportional to $\epsilon(F_\mathrm{intcpt}-F_\mathrm{spec})$, where $\epsilon$ is a proportionality factor that is a function of the antenna's absorptance and front and rear infrared emittance:
\begin{equation}
\vec{F}_\mathrm{diff}=\frac{2}{3}\epsilon(|\vec{F}_\mathrm{intcpt}|-|\vec{F}_\mathrm{spec}|)\vec{n}.
\end{equation}

Using values from Pioneer project documentation\cite{MDR2005,PC202,TCSDR3}, we can calculate $$\epsilon=(1-\alpha_\mathrm{sr})+\alpha_\mathrm{sr}(\epsilon_\mathrm{front}-\epsilon_\mathrm{rear})/(\epsilon_\mathrm{front}+\epsilon_\mathrm{rear})\simeq 0.98,$$ where $\alpha_\mathrm{sr}\simeq 0.21$ is the solar absorptance of the antenna front side, $\epsilon_\mathrm{front}\simeq 0.85$ is its infrared emittance, while $\epsilon_\mathrm{rear}\simeq 0.04$ is the infrared emittance of the antenna rear side. The additional factor of $2/3$ is characteristic of the force due to diffuse emittance by a flat emitter.

The parameters $\alpha\simeq 1$, $-1\le\epsilon\simeq 0.98\le 1$, and $0\le\sigma\le 1$ can be solved for as an orbital solution is obtained. Alternatively, their values can be calculated approximately from project documentation\cite{PC202,TCSDR3}.

\subsubsection{Media effects}

The atmosphere and solar plasma can introduce a significant additional signal propagation delay. One way to deal with this delay is to ignore all data points measured when the spacecraft was less than 5 degrees above the horizon as seen from either the transmitting or the receiving antenna, and when the angular separation between the Sun and the spacecraft was less than 5 degrees. However, it is also possible to estimate these delays.

\begin{itemize}

\item Solar Plasma

Delay due to solar plasma is a function of the electron density in the plasma. Although this can vary significantly as a result of solar activity, the propagation delay $\Delta T$ (in microseconds) can be approximated using the formula\cite{DSMS106}:
\begin{equation}
\Delta T=1.3446\times 10^{-19}\times f^{-2}\times\int N_e~dl,
\label{eq:DeltaT}
\end{equation}
where $f$ is the signal frequency, and $N_e$ is the electron density, which is integrated along the propagation path $l$. For $N_e$, an empirical formula is available\cite{DSMS106}:
\begin{equation}
~~~~~~N_e=2.21\times 10^{14}\left(\frac{r}{R_0}\right)^6+1.55\times10^{12}\left(\frac{r}{R_0}\right)^{-2.3},
\end{equation}
where $R_0=6.96\times 10^8$~m is the solar radius, and $r$ is the distance from the Sun along the propagation path. Although (\ref{eq:DeltaT}) cannot be expressed in closed form, numerical integration can yield a table of values that can be used, along with an interpolation subroutine, to obtain values of $\Delta T$, as a function of the spacecraft-to-Earth distance and the Sun-Earth-spacecraft angle, with sufficient precision.

\item Tropospheric Delay

Chao\cite{SOVERS1994} estimates the delay due to signal propagation through the troposphere using the following formula:
\begin{equation}
R=\frac{1}{\sin E+\frac{A}{\tan E+B}},
\end{equation}
where $R$ is the additional propagation path, $E$ is the elevation angle, and $A=A_\mathrm{dry}+A_\mathrm{wet}$ and $B=B_\mathrm{dry}+B_\mathrm{wet}$ are parameters defined as
\begin{eqnarray}
A_\mathrm{dry}&=&0.00143,\\
A_\mathrm{wet}&=&0.00035,\\
B_\mathrm{dry}&=&0.0445,\\
B_\mathrm{wet}&=&0.017.
\end{eqnarray}

\end{itemize}

\subsubsection{Spacecraft spin}

The radio signal used to communicate with the Pioneer spacecraft is circularly polarized. As the spacecraft itself is spinning, every rotation of the spacecraft effectively adds an extra period to the (received or transmitted) signal. In other words, the spacecraft's spin frequency (1/60th its rate of revolution in units of rpm) must be added to both the received and the transmitted signal.

The sign of the spin correction can be unambiguously determined by comparing archived Doppler data files, some of which were corrected for spin and some of which weren't. This comparison clearly reveals that the spin of the spacecraft must be added to the frequency of both the uplink and the downlink signal with a positive sign.

The spacecraft spin can be directly obtained from telemetry\cite{MDR2005,PC202}. A star sensor, or one of two sun sensors were used to measure the spin period of the spacecraft on board, and this value was telemetered to the Earth using a clock resolution of 1/8192 seconds. This information was sent to the Earth in the form of three telemetry words %named $C_{405}$, $C_{406}$, and $C_{407}$
in the telemetry data stream%. The spin period can be reconstructed using the expression
\cite{MDR2005}.

The sun sensor on the spacecraft was operational only up to a distance of ~35 AU from the Sun. On Pioneer 10, the star sensor failed after Jupiter encounter. As a result, once the spacecraft passed beyond 35 AU, the spin period could no longer be determined on board. Readings from the Imaging Photo-Polarimeter instrument were used to occasionally derive further estimates of the spin rate.

\subsubsection{Maneuvers}

After planetary encounters, the Pioneer spacecraft no longer executed trajectory correction maneuvers. However, they still periodically performed precession maneuvers, designed to ensure that the spacecraft remains pointed in the direction of the Earth with sufficient accuracy to maintain reliable two-way radio communication.

These maneuvers involved the firing of thrusters on opposite sides of the spacecraft, one in the fore, the other in the aft direction, for equal duration. In principle, such a maneuver imparts no net change in momentum on the spacecraft. In practice, uncertainties in thruster alignment and firing duration result in a small net change in the spacecraft velocity in a random direction. As the spacecraft spins, random changes in momentum in a direction perpendicular to the spin axis cancel out over time, only increasing the uncertainty in the spacecraft's lateral position in the sky slightly. However, the firings may increase or decrease the spacecraft's velocity along its spin axis, and this must be taken into account.

The magnitude of this velocity change is essentially random; it cannot be predicted. However, the exact time of precession maneuvers is known from the telemetry data stream.

Specifically, the telemetry data stream provides thruster pulse counts\cite{MDR2005,PC202}; 6-bit counters for each thruster that are incremented every time the thruster in question is fired.

We were able to compile this information from the telemetry into a list of times when precession maneuvers occurred. This list is directly imported into our orbit determination code if an estimate of maneuvers is requested.

\subsubsection{Heat}

The study of on-board generated heat and its effects on the orbit of Pioneer 10 and 11 are a major focus of our on-going investigation\cite{MDR2005,MDR2006,MEXPROC2007}. Our orbit determination code has the ability to deal with heat sources and calculate their contribution to acceleration. This capability will be used with the newly compiled Doppler data set; results will be reported as they become available.

\subsubsection{Radio beam recoil force}

The spacecraft's radio beam also carries momentum that corresponds to the approximately 8~W of transmitted power. The spacecraft's HGA is always aimed at the approximate direction of the Earth, deviating from that orientation by no more than a couple of degrees. However, it has been argued\cite{LKS2003} that not all of the radio power emitted by the transmitter hits the reflecting dish of the HGA; some of it, typically 10\%, would miss the antenna and would be radiated in the opposite direction, at an approximately 45$^\circ$ angle. This means that the radio beam will translate into momentum by an approximate efficiency of $\sim 83\%$\cite{LKS2003}.

The actual power of the transmitter was always assumed to be constant, but telemetry suggests otherwise: the transmitter power is directly measured, and it varied significantly\cite{MDR2005,MDR2006}.

\subsubsection{Solar system data}

Rather than integrating the equations of motion for all solar system bodies, we use the planetary ephemerides published by JPL. Specifically, we are using JPL's CSPICE library\footnote{See {\tt http://naif.jpl.nasa.gov/}.} to obtain high accuracy planetary positions. To calculate the orbit of the Pioneer spacecraft, we utilize the state vectors of the Sun and the eight major planets from the DE-414 ephemerides.% All planetary systems are treated as pointlike bodies, using the planetary system's barycenter state vector.

Except for cases when the spacecraft are near Jupiter or Saturn, we ignore the effects of individual moons. We also ignore smaller bodies such as Pluto, Ceres, Charon, or the larger Kuiper Belt objects.

The CSPICE library is also used to obtain accurate estimates of planetary masses, multiplied by the gravitational constant. Our work therefore does not depend on the accurate value of the gravitational constant, and all calculations are based on data sets with a consistent origin.

We also used the CSPICE library to obtain accurate state vectors of the participating stations of the Deep Space Network. The library allows us to obtain momentary positions and velocities. However, a typical Doppler measurement has a duration ranging from several seconds to several minutes, during which the station moves significantly in a solar system barycentric frame of reference, and its velocity changes as well. In a geocentric reference frame, the station moves along a circular arc in the plane of the equator. For an arc spanning an angle $\phi$ with radius $r$, the center of gravity is located at angle $\phi/2$ at a distance of $[(\sin\phi/2)/(\phi/2)]r$ ($\phi$ measured in radians) from the origin. This expression was used to correctly average the station position due to the Earth's rotation.

The right ascension and declination of Jupiter's and Saturn's north pole are provided in the appropriate CSPICE kernels, along with spherical harmonic coefficients ($J_2$ through $J_8$).

In the vicinity of the Earth, Jupiter, and Saturn, the gravitational influences of individual moons must be considered. %Satellites with sufficient mass to appreciably alter the gravitational field in the vicinity of the corresponding planet are listed in Table~\ref{tab:sats}.
Systems of satellites need not be considered when the spacecraft is far from the planetary system. A distance of $10^{11}$~m ($\sim$ 2/3 AU) works as a sensible cut-off.

Jovian and Saturnian satellite ephemerides are obtained from the CSPICE kernel files JUP204 (Galilean satellites, DE-405), JUP256 (minor satellites, DE-405), and SAT252 (Saturnian satellites, DE-414.)

Station position and velocity information (in the Earth-centered ITRF93 reference frame) for all operating DSN stations are readily available in CSPICE kernel files. It is somewhat harder to obtain station positions for DSN stations that have been decommissioned, or DSN stations that were relocated.

Specifically, position and velocity information for DSS-11, DSS-41, DSS-44, DSS-51, and DSS-62 were provided by Folkner\footnote{Private communication.}, along with position information for DSS-12 and DSS-61 before their relocation in 1978-79. An ambiguity concerning the exact date of the relocation of DSS-12 was resolved by Ref.~\refcite{UPDOWN}.% This information was compiled into an additional CSPICE kernel file that was used in conjunction with \texttt{earthstns\_itrf93\_050714}.

\subsubsection{Orbit data files}

Actual measurements of the Pioneer spacecraft's radio frequency are stored in files called ODFs (Orbit Data Files.) These files have a simple binary format encoding the following pieces of information:
\begin{itemize}
\item Time of reception
\item Count period
\item Reference frequency
\item Transmitting station identifier
\item Receiving station identifier
\item Doppler observable.
\end{itemize}

In principle, the Doppler observable can be reconstructed from the first five of these values: from the time of reception and the known orbit of the spacecraft, taking light trip time into account, the time of transmission on board the spacecraft can be calculated, and from that, the time of transmission at the transmitting station can be obtained. The relative velocities of these three locations, combined with the reference frequency, can be used to obtain the received frequency; the difference between this and the reference frequency is the Doppler reading.

One complicating factor is that the actual frequency of transmission might not have been the reference frequency; indeed, the reference frequency itself might have been changing in time as well, i.e., in many cases these frequencies were ``ramped''. The ODFs contain ramp data records that can be used to determine the precise transmission and reference frequencies at specific DSN stations at any given time. The absence of a ramp record typically manifests itself as an offset of several kHz between the calculated and observed frequency; Doppler data records with missing ramp information can thus be easily identified and rejected.

The ODFs also had other correctable issues, such as incorrect encoding of frequencies (e.g., marking an X-band frequency as a Ku-band frequency) that were easily fixed.

A far more insidious problem is related to spin correction. Often, ODFs are adjusted to account for the spacacraft's spin, but this is not always the case. Notably, the 11.5 year Pioneer data set contained records both with and without spin correction; data points prior to July 18, 1990 had spin correction applied, whereas later data points were uncorrected\footnote{Private communication from P. Tortora.}.

\section{Results}
\label{sec:results}

The considerations presented so far were used to build an orbit determination program that took into account solar system ephemerides published by JPL and utilized through the CSPICE library, DSN station locations on the Earth, again calculated using the CSPICE library and averaged for the duration of Doppler intervals, Doppler observations obtained from ODFs, spacecraft spin data obtained from telemetry, the spacecraft's trajectory, light trip times including the Shapiro delay, a flat disk solar pressure model, and optionally, additional telemetry/design values used to estimate thrust due to heat and the radio beam recoil force. The goal of this application was to minimize the sum of the squares of differences between computed and observed Doppler; the minimization algorithm used was the nonlinear least squares algorithm with approximate Jacobian from the standard MINPACK library\footnote{See {\tt http://www.netlib.org/minpack/}.}. We weighted individual data points using the Doppler duration, with the assumption that the standard deviation of the data is the inverse of that duration; for instance, if the Doppler count was 1000 seconds, we assume that the data point is accurate to 1 ms within 1 standard deviation.

With all the tools in place, the first order of business was to use these to confirm the earlier results regarding the Pioneer anomaly.

\subsection{Confirmation of the Pioneer anomaly}

To confirm previous results regarding the anomalous acceleration of Pioneers 10 and 11, we used ODFs provided by JPL. These ODFs contained verified and filtered records spanning the time period 1986-1998 (Pioneer 10) and 1986--1990 (Pioneer 11). These were the exact same data records examined by JPL in 2002\cite{JPL2002}.

\begin{figure}
\begin{minipage}[b]{.49\linewidth}
\includegraphics[width=\linewidth]{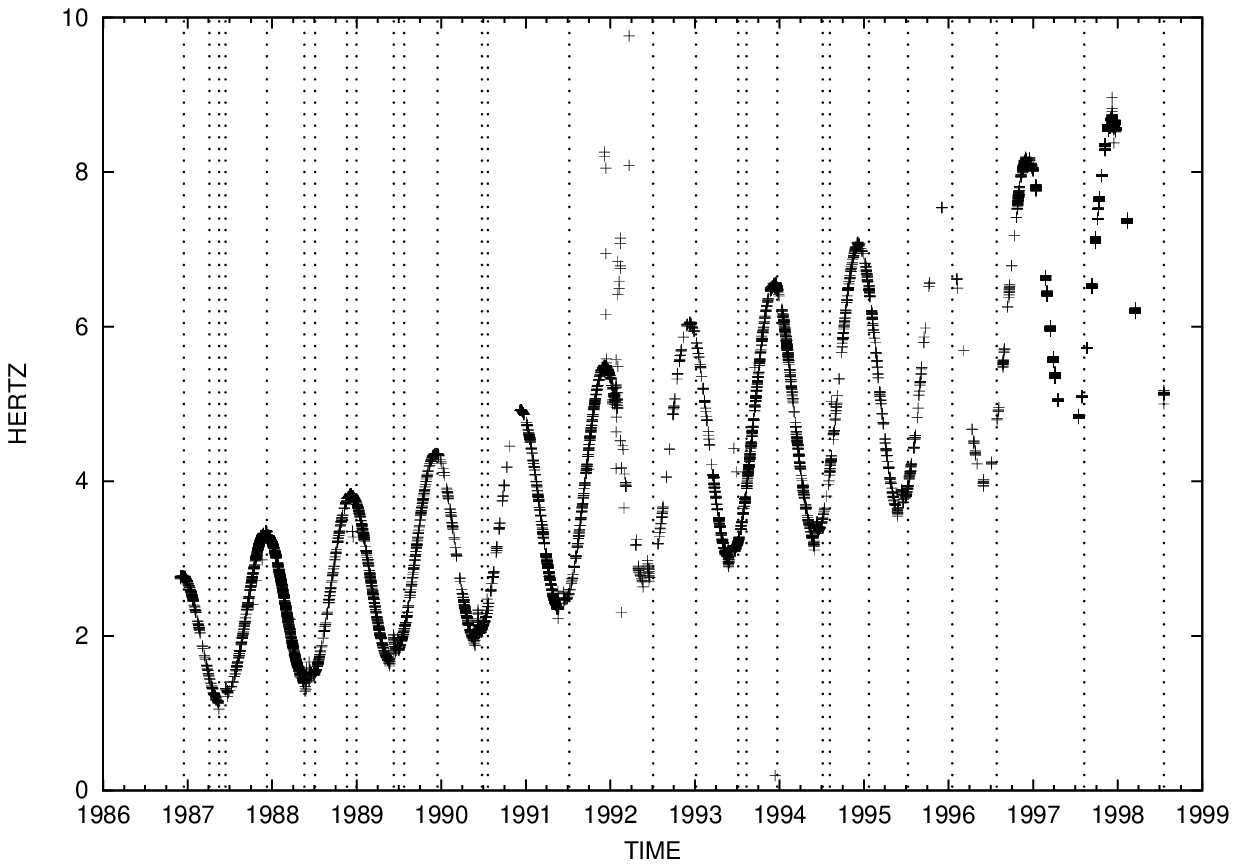}
\end{minipage}
\begin{minipage}[b]{.49\linewidth}
\includegraphics[width=\linewidth]{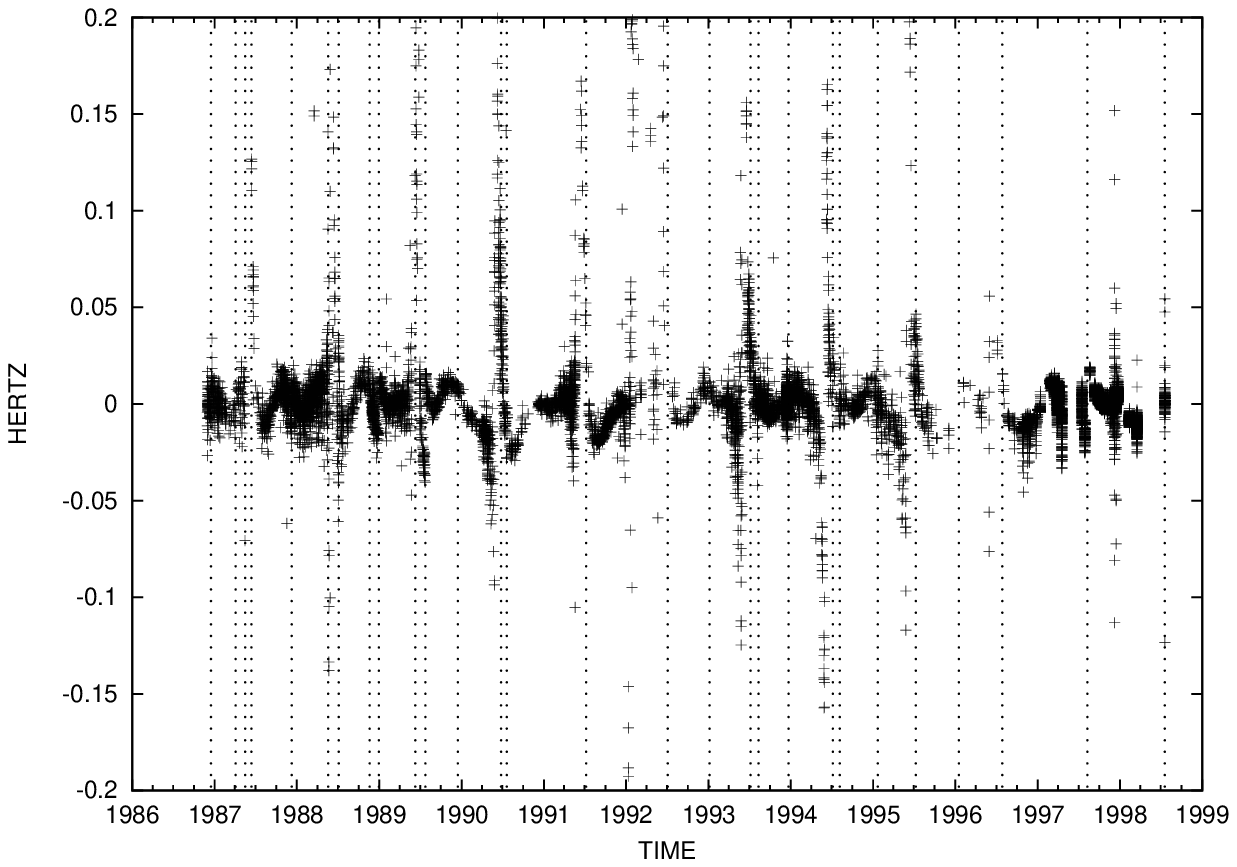}
\end{minipage}
\newline
\begin{minipage}[b]{.49\linewidth}
\includegraphics[width=\linewidth]{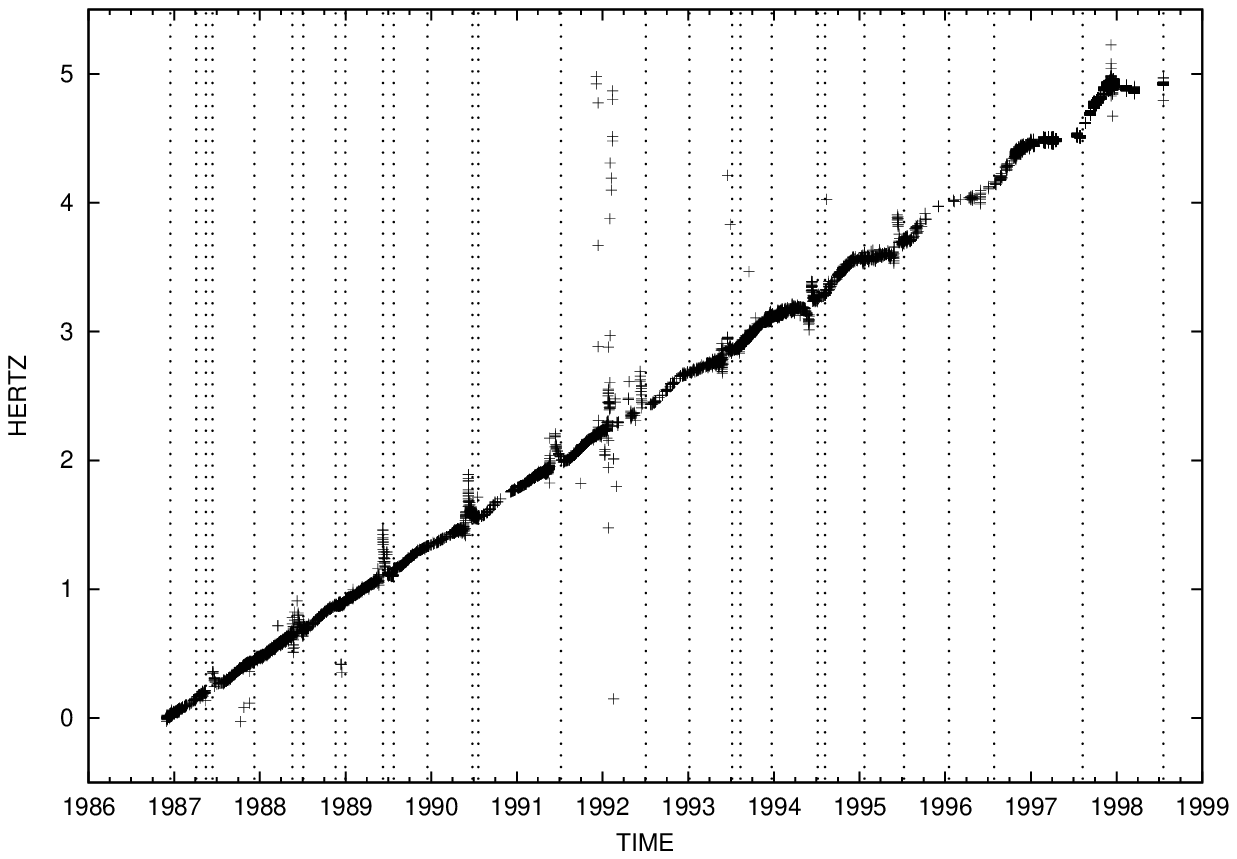}
\end{minipage}
\begin{minipage}[b]{.49\linewidth}
\includegraphics[width=\linewidth]{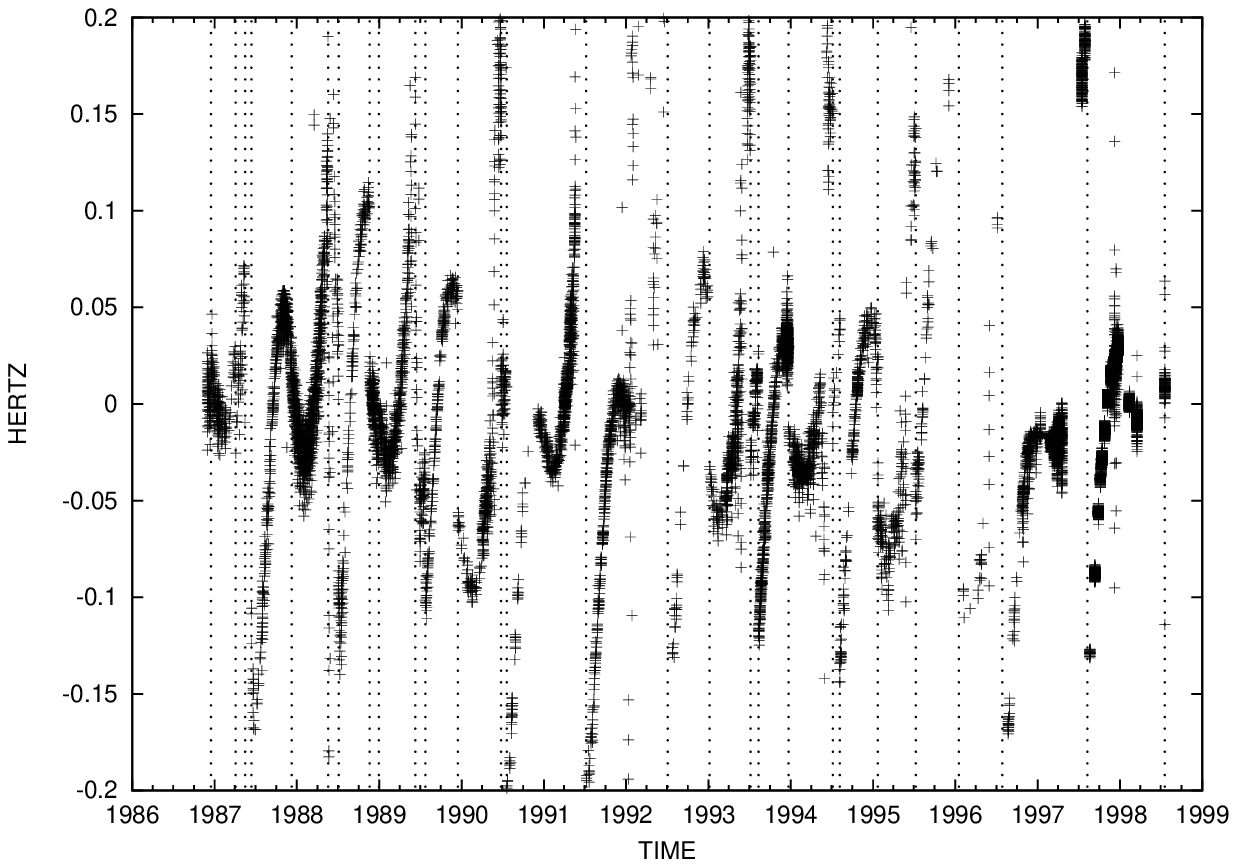}
\end{minipage}
\caption{Pioneer 10 Doppler residuals. Vertical axis is residual frequency in Hz; horizontal axis is time (in years). Dotted vertical lines correspond to maneuvers. Top left: residuals prior to fitting. Top right: residuals after a best fit of the initial state vector, the anomalous acceleration and 28 maneuvers.
Bottom left: residuals after the anomalous acceleration has been removed from the best fit shown in top left. Bottom right: residuals after a best fit of the initial state vector and 28 maneuvers, without anomalous acceleration.}
\label{fig:p10}
\end{figure}

\begin{figure}
\begin{minipage}[b]{.49\linewidth}
\includegraphics[width=\linewidth]{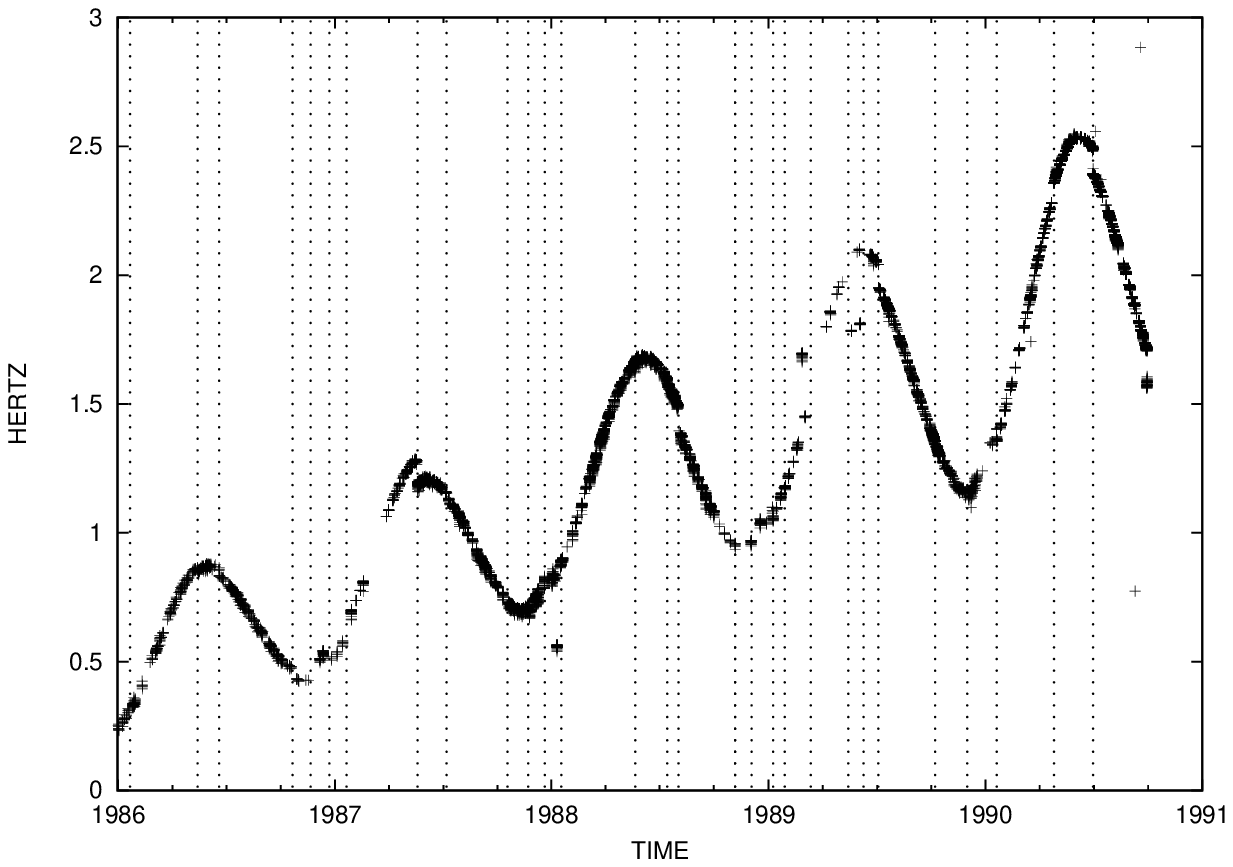}
\end{minipage}
\begin{minipage}[b]{.49\linewidth}
\includegraphics[width=\linewidth]{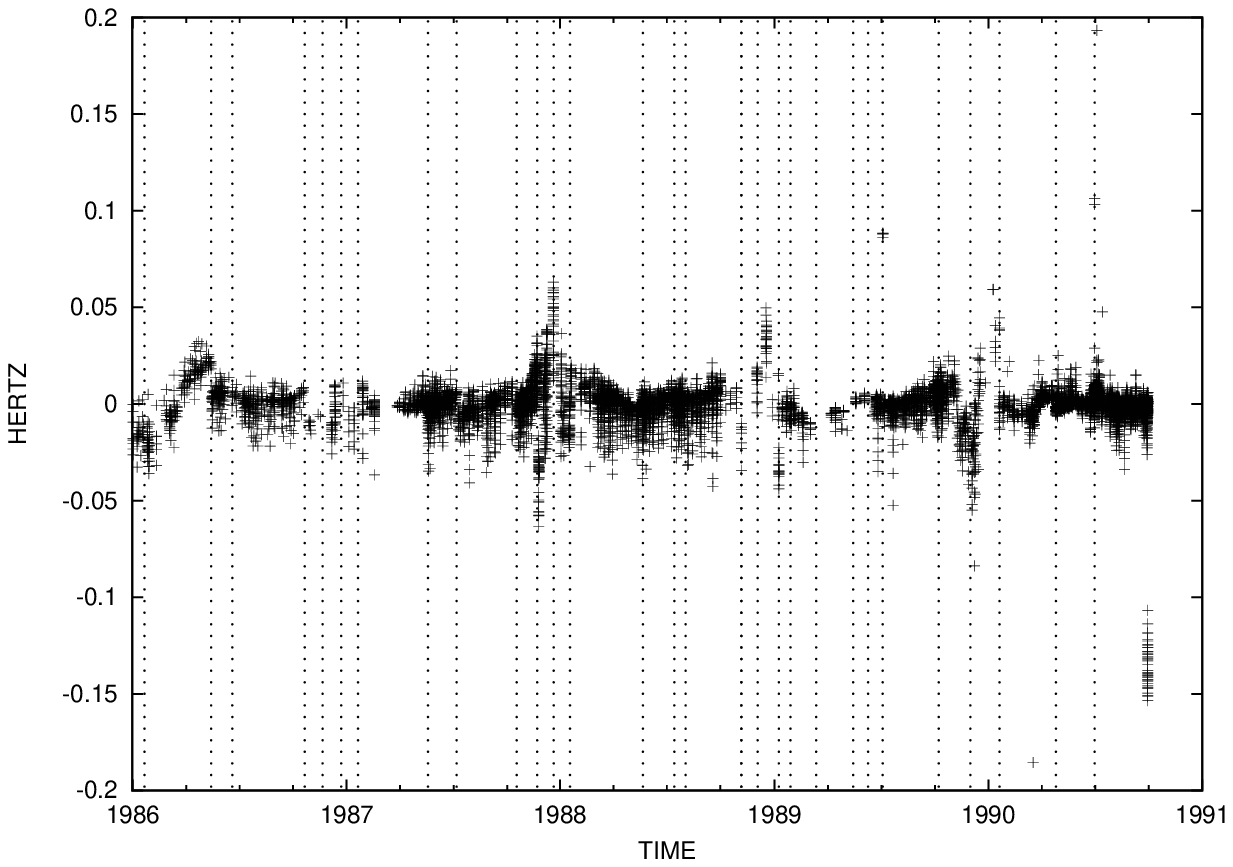}
\end{minipage}
\newline
\begin{minipage}[b]{.49\linewidth}
\includegraphics[width=\linewidth]{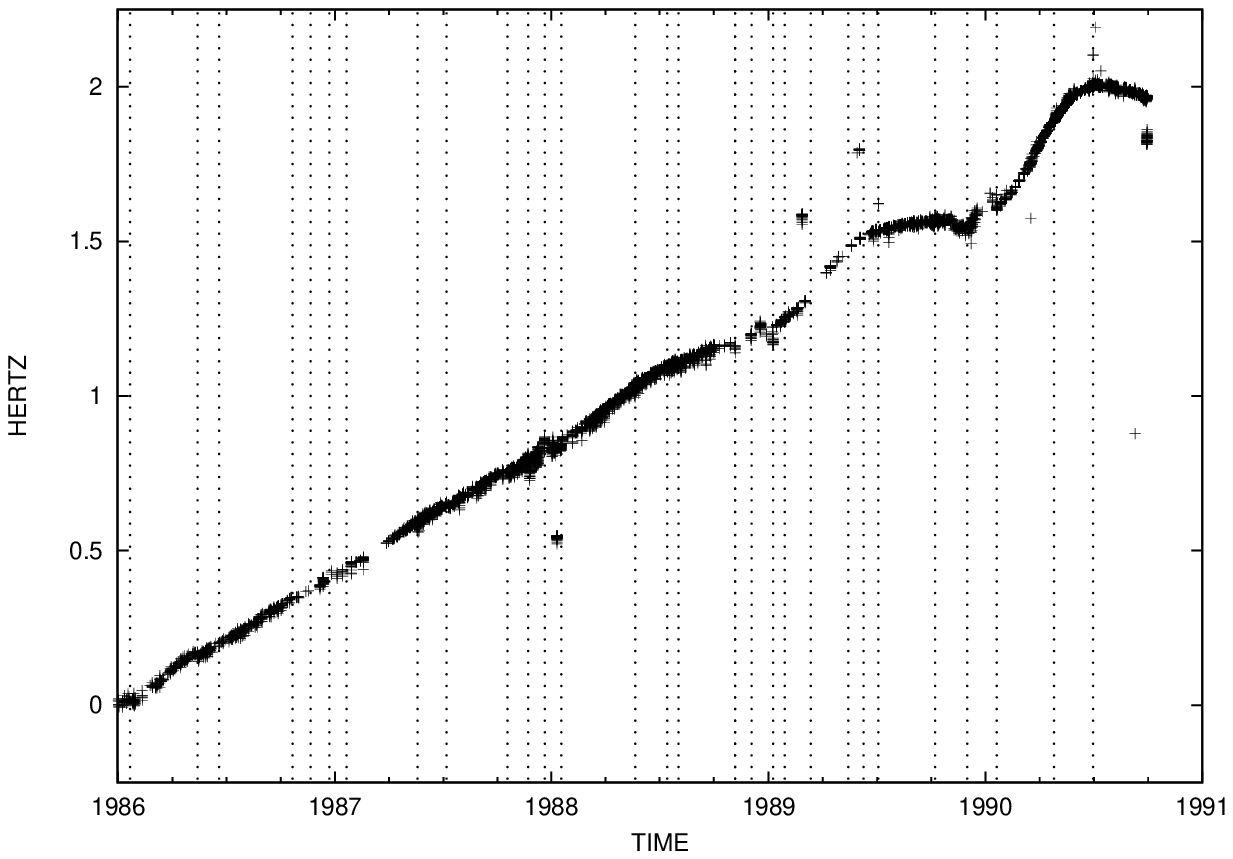}
\end{minipage}
\begin{minipage}[b]{.49\linewidth}
\includegraphics[width=\linewidth]{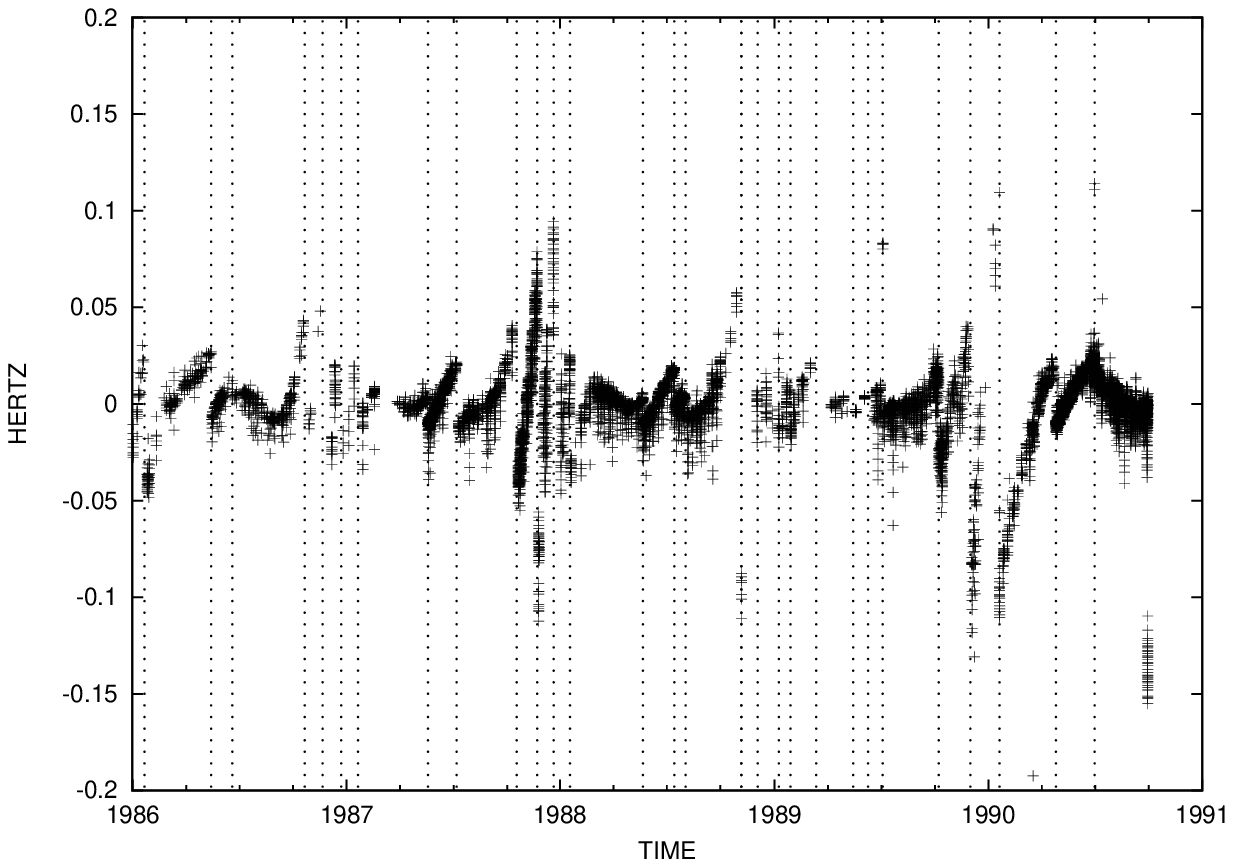}
\end{minipage}
\caption{Pioneer 11 Doppler residuals. Vertical axis is residual frequency in Hz; horizontal axis is time (in years). Dotted vertical lines correspond to maneuvers. Top left: residuals prior to fitting. Top right: residuals after a best fit of the initial state vector, the anomalous acceleration and 30 maneuvers. Bottom left: residuals after the anomalous acceleration has been removed from the best fit shown in top left. Bottom right: residuals after a best fit of the initial state vector and 30 maneuvers, without anomalous acceleration.}
\label{fig:p11}
\end{figure}

For the initial state vector, we used the approximate Pioneer ephemerides from JPL HORIZONS. This produced a good fit already, with the Doppler residual for most data points not exceeding a few Hz (Figure~\ref{fig:p10} top left and Figure~\ref{fig:p11} top left.)

\subsection{Pioneer 10 results}

For Pioneer 10, we used JD (Julian Day) 2446765 through JD~2451016 as the period of investigation; the data set contained 19758 Doppler data points in total. About one third of the Doppler data points were obvious outliers (yielding residuals of up to several MHz). Most of these data points were in the last two years of the mission, corresponding to changes in DSN station configuration, and therefore, it is possible that these ODF records are misinterpreted by our code. Nevertheless, after the removal of outliers, a sufficient number of good data points remains for analysis. The time period included 28 maneuvers. We used an incremental filtering strategy to process this data set. We used a nominal spacecraft mass of 250~kg, corresponding to a loss of one quarter of the propellant on board, when calculating nongravitational forces, such as solar pressure. In addition to filtering outliers, we also set a minimum horizon angle of $5^\circ$ and a minimum angular separation angle of $5^\circ$ between the spacecraft and the Sun. Our initial run, prior to parameter fitting, yielded 13525 data points.

To fit the data, we followed an incremental strategy. First, we estimated the initial state vector, maneuvers, and the constant sunward acceleration using every 20th record in the data set. This allowed for a rapid process to find good initial estimates for these parameters, allowing us to iteratively reduce the filter bandwidth; eventually, we were able to narrow the bandwith to 100~mHz, rejecting data points that yielded a Doppler residual in excess of 50~mHz. At the end of this stage, we were left with 13119 Doppler data points.

Finally, we fitted these data points without any additional filtering or sampling. We obtained a result of $(9.03\pm 0.13)\times 10^{-10}~\mathrm{m}/\mathrm{s}^2$. Combined with error terms from Ref.~\refcite{JPL2002} (but excluding terms related to thermal modeling, which is the subject of our on-going effort), amounting to $\pm 0.85\times 10^{-10}$~m/s$^2$, we obtain
\begin{equation}
a_\mathrm{P10}=(9.03\pm 0.86)\times 10^{-10}~\mathrm{m}/\mathrm{s}^2,
\end{equation}
with a root mean square (rms) Doppler residual of 10.1~mHz (Figure~\ref{fig:p10} top right).

Figure~\ref{fig:p10} (bottom left) shows the Doppler residuals using the fitted value of the initial state vector and maneuvers, but with $a_P$ set to 0. The result is an unambiguous drift, another clear indication of the anomalous acceleration.

Is it possible to fit the data with $a_P=0$? Clearly, much of the anomalous behavior can be absorbed into the 28 maneuvers that took place during this time period. However, one would expect a worse fit in this case. Indeed, this is what we found; after fitting 13111 data points by varying only the initial state vector and maneuvers, but without an anomalous acceleration component, the rms residual was 43.8~mHz (Figure~\ref{fig:p10} bottom right). This clearly indicates that the constant acceleration term offers a statistically significant, superior fit to the Doppler data.

\subsection{Pioneer 11 results}

The results for Pioneer 11 are similar to the Pioneer 10 results. The Doppler data, from JD~2446431 to JD~2448256, covered approximately five years, with a total of 11423 data points, containing only a small number of outliers. The nominal spacecraft mass used was 232~kg, to account for the increased use of propellant (three quarters of the propellant on board) during Pioneer 11's trajectory correction maneuvers. For Pioneer 11, a shorter data span was available, but it contained more data points; there were also more maneuvers performed during the study period. Our initial filtering yielded 11230 data points (Figure~\ref{fig:p11} top left).

Following an incremental editing strategy just like in the case of Pioneer 10, we were able to fit the initial state vector, 30 maneuvers, and the anomalous acceleration. With 11080 data points remaining in the data set after filtering, we obtained $(8.21\pm 0.66)\times 10^{-10}$~m/s$^2$, which, combined with error terms from Ref.~\refcite{JPL2002}, yields
\begin{equation}
a_\mathrm{P11}=(8.21\pm 1.07)\times 10^{-10}~\mathrm{m}/\mathrm{s}^2,
\end{equation}
with an rms residual of 8.2~mHz. The substantially larger formal error can be attributed to the shorter timespan covered by the data set, and more frequent maneuvers (Figure~\ref{fig:p11} top right).

Figure~\ref{fig:p11} (bottom left) shows the computed Doppler residual using the fitted initial state vector and maneuvers, but with $a_P$ set to zero. Again, the anomalous velocity drift is clearly indicated in the data.

As in the case of Pioneer 10, we also attempted to fit the Doppler data using the initial state vector and maneuvers only, with the anomalous acceleration set to zero. The result, shown in Figure~\ref{fig:p11} (bottom right), is inferior, though not as clearly so as was the case for Pioneer 10. The rms residual remains a modest 12.4~mHz on 11070 data points, due to the fact that the shorter timespan and greater frequency of maneuvers allowed a tighter fit even without $a_P$.

\subsection{Spectral Analysis}

How good are the results that we obtained so far? As can be seen from the unfitted residuals (Figure~\ref{fig:p10} top left and Figure~\ref{fig:p11} top left), any mismodeling is bound to produce a clear annual signature, due to the orbital motion and rotation of the Earth where the DSN antennae are located.

\begin{figure}
\begin{minipage}[b]{.49\linewidth}
\includegraphics[width=\linewidth]{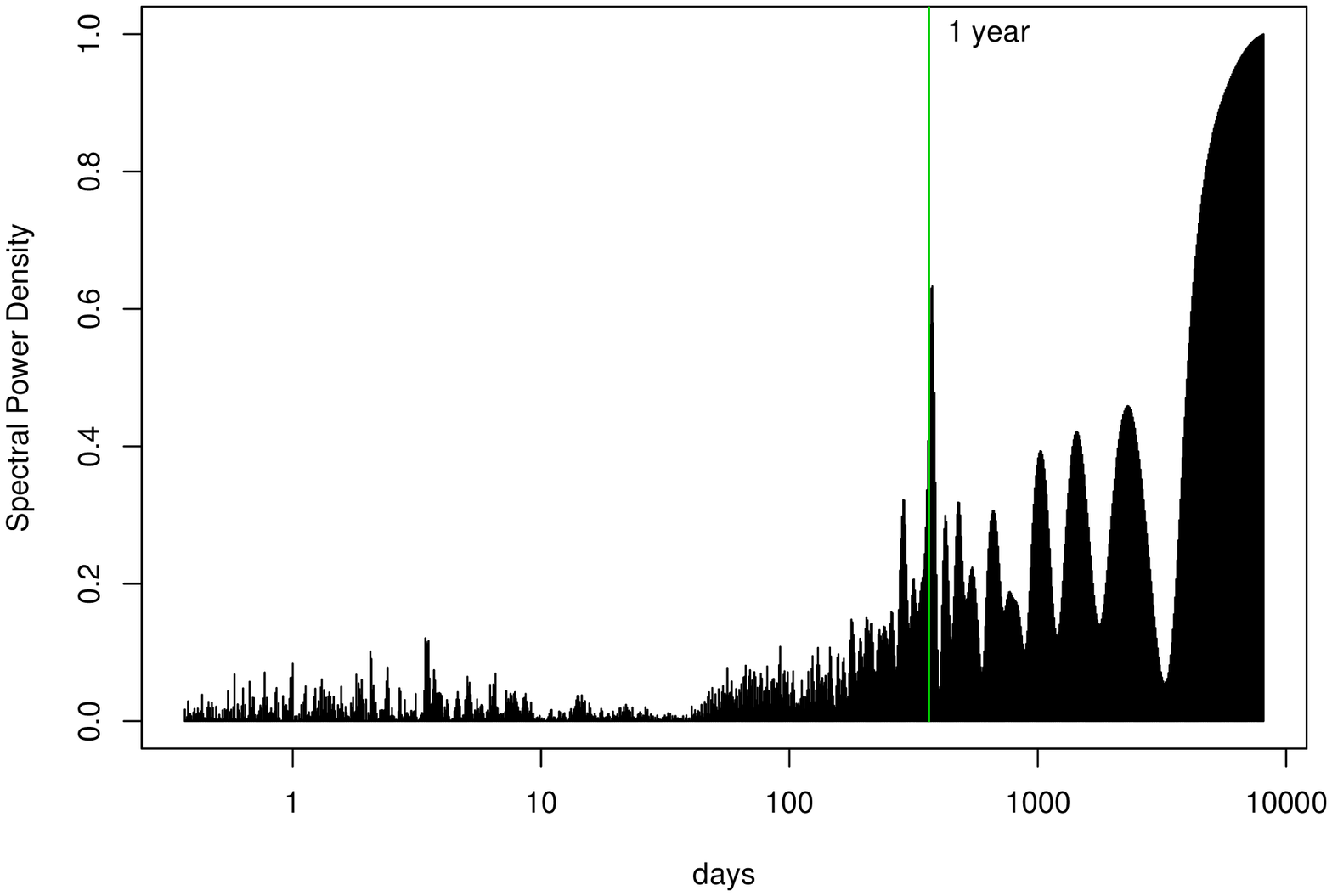}
\end{minipage}
\begin{minipage}[b]{.49\linewidth}
\includegraphics[width=\linewidth]{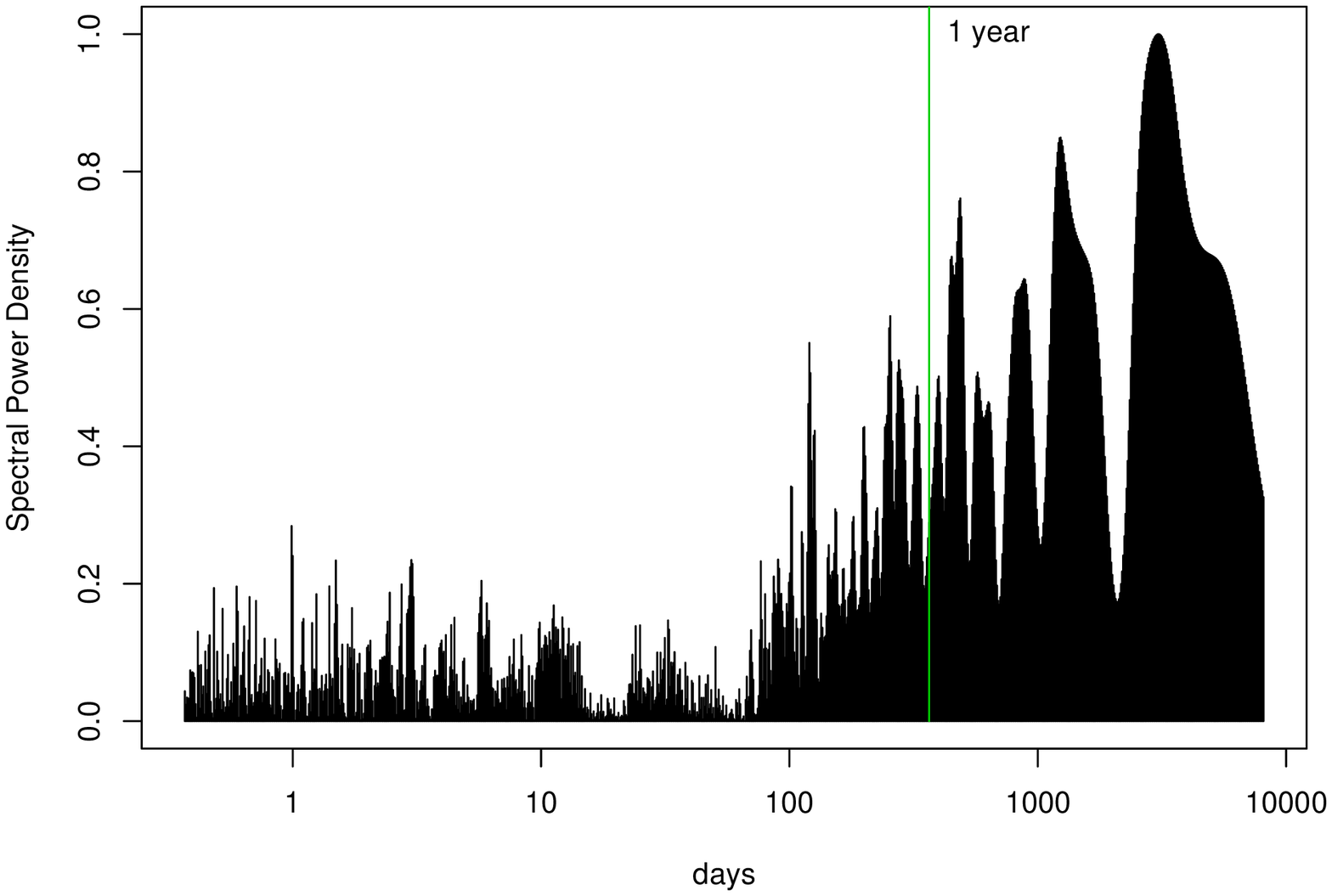}
\end{minipage}
\caption{Pioneer 10 residuals normalized spectral density. Left: residuals prior to fitting; a strong annual signal (position indicated by vertical green line) is clearly seen. Right: residuals after the fit depicted in Figure~\ref{fig:p10} (top right). The annual signature has vanished, and only a small peak is visible at a period of 1 day, indicating a small diurnal mismodeling.}
\label{fig:p10_sp}
\end{figure}

\begin{figure}
\begin{minipage}[b]{.49\linewidth}
\includegraphics[width=\linewidth]{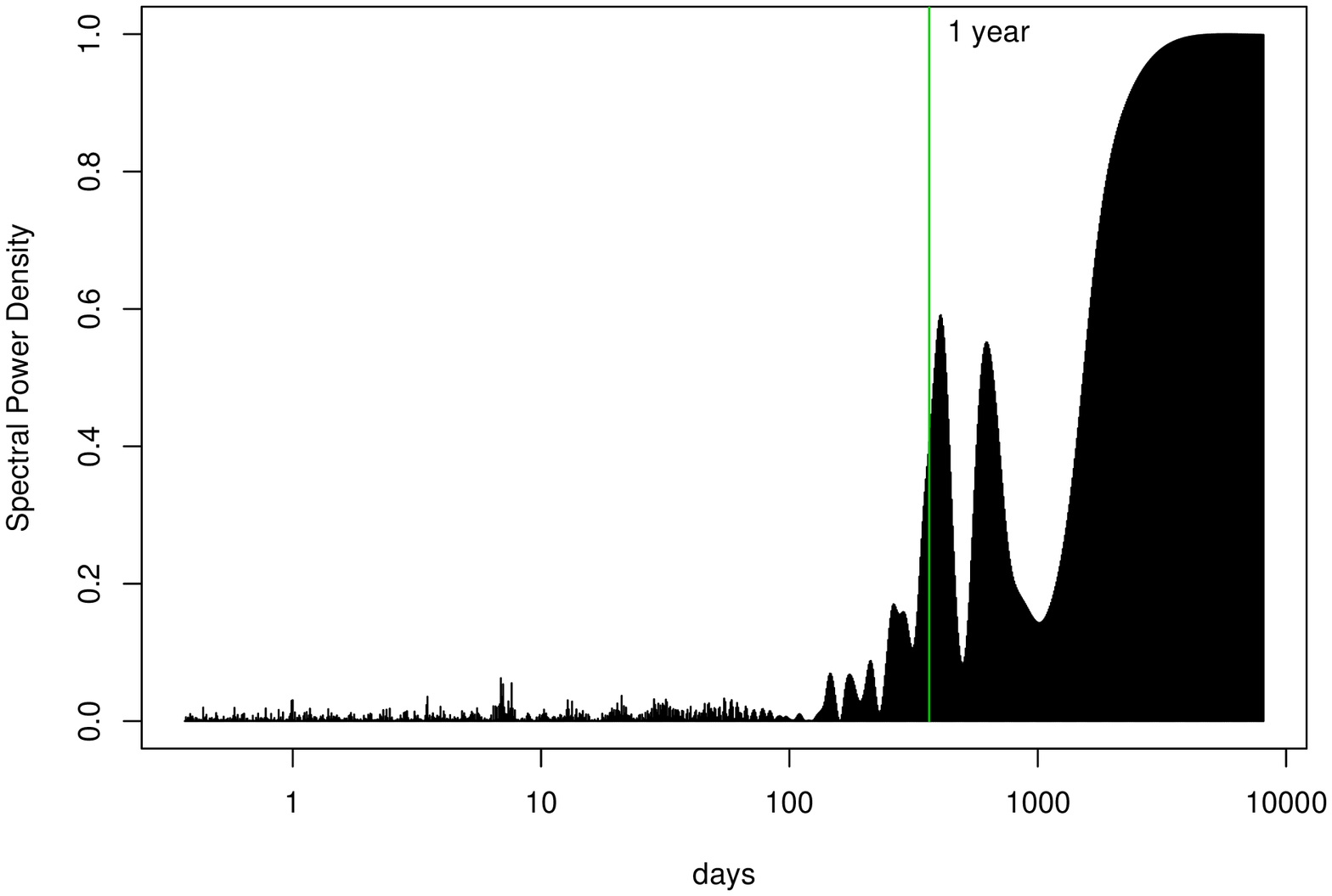}
\end{minipage}
\begin{minipage}[b]{.49\linewidth}
\includegraphics[width=\linewidth]{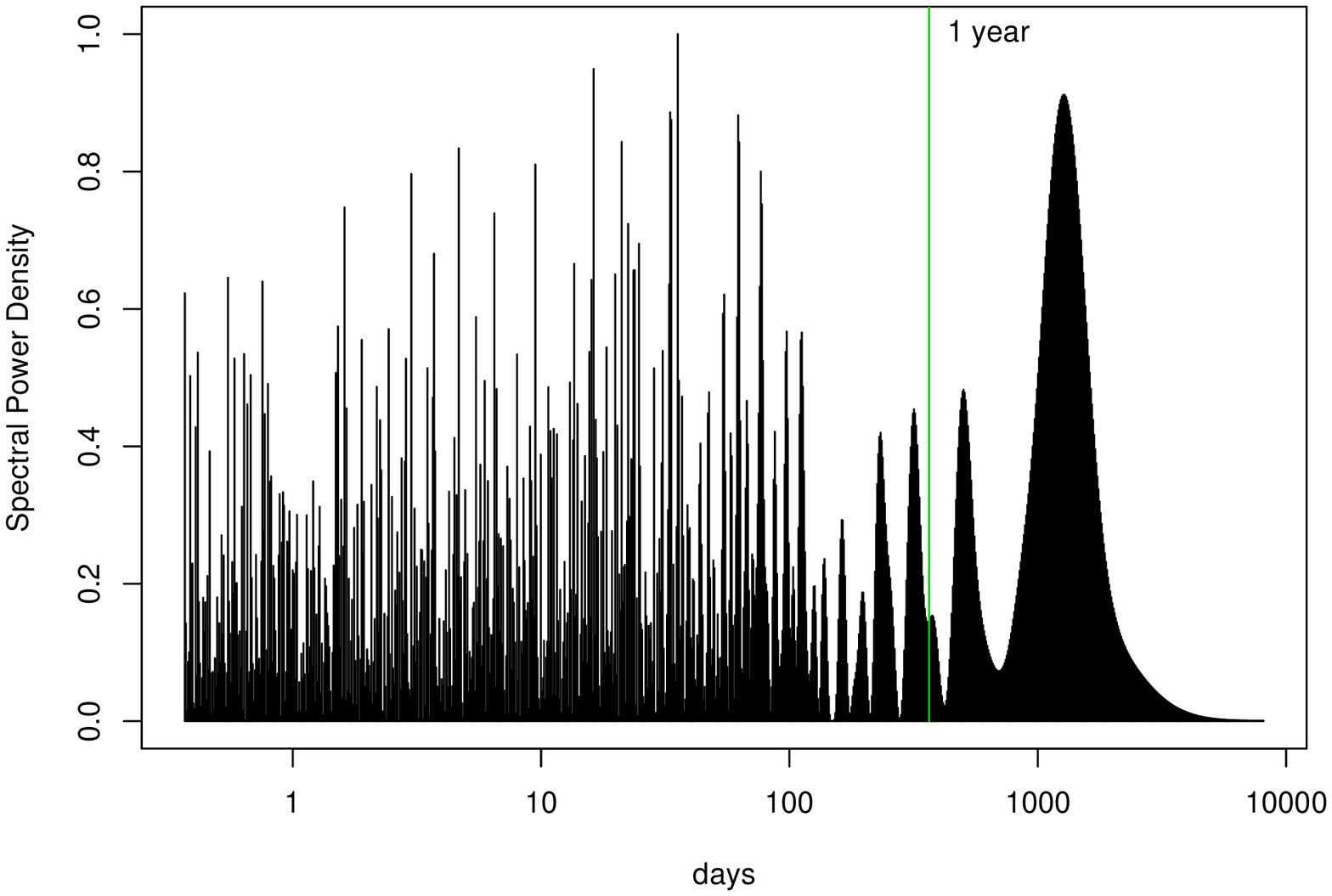}
\end{minipage}
\caption{Pioneer 11 residuals normalized spectral density. Left: residuals prior to fitting; a strong peak is visible at a period of $\sim$370 days, which is an annual signal the period of which is skewed by the spacecraft's proper motion. Right: residuals after the fit depicted in Figure~\ref{fig:p11} (top right). No obvious peaks are visible that would indicate a diurnal or annual mismodeling.}
\label{fig:p11_sp}
\end{figure}

Such signatures must be observable in the power spectrum of Doppler residuals. Because the sampling intervals are not uniform (there are dropouts and blackouts in the Doppler data stream), traditional FFT methods are not readily applicable to our residuals; however, the residuals can be readily analyzed using Lomb--Scargle methods\cite{NUMRC,GLYNN2006}.

The results for the unfitted data sets, shown in Figure~\ref{fig:p10_sp} (left) and Figure~\ref{fig:p11_sp} (left) clearly indicate the presence of an annual signal (with a period slightly longer than one year due to the spacecrafts' lateral motion). A small peak is also visible at a period of 1 day, though it is not significant compared to other peaks.

After the fit, the annual terms have all but vanished from the power spectrum (Figure~\ref{fig:p10_sp} right, and Figure~\ref{fig:p11_sp} right). For Pioneer 10, a noticeable peak is present at a period of 1 day, indicating a possible diurnal mismodeling; no such peak can be observed in the Pioneer 11 post-fit power spectrum.

\subsection{Jerk term}

A recurring question concerning the Pioneer anomaly is its constancy. To test the extent to which the anomalous acceleration is constant in time, we implemented the ability to estimate a secondary acceleration, i.e., ``jerk'' term in our orbital solution.

For Pioneer 10, we obtained

\begin{eqnarray}
a_\mathrm{P10}&=&(10.96\pm 0.89)\times 10^{-10}~\mathrm{m}/\mathrm{s}^2,\\
\dot{a}_\mathrm{P10}&=&(-0.21\pm 0.04)\times 10^{-10}~\mathrm{m}/\mathrm{s}^2/\mathrm{year},
\end{eqnarray}
with an rms residual of $9.8$ mHz on 13130 data points.

For Pioneer 11, the result was
\begin{eqnarray}
a_\mathrm{P11}&=&(9.40\pm 1.12)\times 10^{-10}~\mathrm{m}/\mathrm{s}^2,\\
\dot{a}_\mathrm{P11}&=&(-0.34\pm 0.12)\times 10^{-10}~\mathrm{m}/\mathrm{s}^2/\mathrm{year},
\end{eqnarray}
with an rms residual of $8.1$ mHz on 10995 data points.

We note that this jerk term is consistent with the expected temporal variation of a recoil force due to heat generated on board and emitted anisotropically.

\section{Plans and conclusions}
\label{sec:plans}

The results presented in this work confirm the existence of the Pioneer anomaly. Doppler observations from the twin Pioneer spacecraft only match calculated orbits if we assume the presence of a small, approximately constant anomalous acceleration in the sunward direction.

We have demonstrated our ability to reproduce results using independently developed orbit determination software with sufficient accuracy. In addition to the Doppler data files, our software has the ability to process telemetry data obtained from the spacecraft. Specifically, telemetry records are used to establish the spacecraft's spin and the timing of maneuvers. The telemetry records can also be used to estimate thermal emissions from the spacecraft, which is the subject of our on-going study.

In addition to confirming the numerical value of the anomalous acceleration, we also considered a secondary acceleration (``jerk'') term. We demonstrated that a moderate jerk term is consistent with the Doppler data, and therefore, an anomalous acceleration that is a slowly changing function of time cannot be excluded at present.

Our modeling can be improved. We can presently model planetary flybys with limited precision; the maximum Doppler residual during flybys of Jupiter and Saturn is as high as a few Hz (indicating a velocity mismodeling of several 10 cm/s) leaving room for improvement. We also plan to make our tropospheric model more sophisticated; even in the absence of historical weather records, Markwardt suggests that seasonal weather data can be used\footnote{Private communication.} to achieve good modeling accuracy. We may also model signal propagation delays due to charged particles in the Earth's ionosphere.

Our main effort, however, is aimed at developing robust orbit estimates that incorporate a telemetry-driven time dependent model of on-board generated thermal recoil forces. The goal is to determine unambiguously the extent to which thermal forces may be responsible for the anomalous acceleration of the Pioneer spacecraft.

\section*{Acknowledgments}

I would like to thank Dr. Slava Turyshev of JPL for his support, advice, and encouragement; and Larry Kellogg, formerly of NASA Ames (now retired), who singlehandedly saved the telemetry records of Pioneer 10 and 11.

\bibliography{refs}
\bibliographystyle{unsrt}

\end{document}